\documentclass[preprint,authoryear,12pt]{elsarticle}

\usepackage{epsfig}

\usepackage{amssymb}

\usepackage[ps2pdf,%
a4paper=true,%
breaklinks=true,%
colorlinks=true,%
pdfauthor={First Author et al.},%
pdftitle={Template for manuscripts in Advances in Space Research}%
]{hyperref}

\journal{Advances in Space Research}

\begin{document}

%%%%%%%%%%%%%%%%%%%%%%%%%%%%%%%%%%%%%%%%%%%%%%%%%%%%%%%%%%%%%%%%%%%%%%%%%%%%%
%% Frontmatter
\begin{frontmatter}


% Use the tnoteref command within \title and fnref within \author or
%\address for footnotes;
% use the corref command within \author for corresponding author
 % footnotes;
% \author{Name\corref{cor1}\fnref{label2}}
% \address{Address\fnref{label3}}

\title{On ultra-high energy cosmic rays: origin in AGN jets and transport
in expanding Universe}
%\tnotetext[footnote1]{This template can be used for all publications in Advances in Space Research.}

% Use optional labels to link authors explicitly to addresses:

\author{Vladimir Ptuskin\corref{cor}, Svetlana Rogovaya and Vladimir Zirakashvili}
\address{Pushkov Institute of Terrestrial Magnetism, Ionosphere
and Radio Wave Propagation (IZMIRAN), Troitsk, Moscow region
142190, Russia}
\cortext[Vladimir Ptuskin]{}
%\fntext[footnote2]{Additional information regarding the
%corresponding author}
\ead{vptuskin@izmiran.ru;rogovaya@izmiran.ru;zirak@izmiran.ru}

% Url can be given like this:
% \ead[url]{http://www.elsevier.com/wps/find/authorsview.authors/latex}

%\ead{more@email.addresses}

%\author{More Authors\fnref{footnote4}}
%\address{Address of the co-authors}
%\fntext[footnote4]{Additional information about the co-authors}
%\ead{more@email.addresses}

\begin{abstract}

The cosmic ray source spectrum produced by AGN (Active Galactic
Nucleus) jets is calculated. A distinctive feature of these
calculations is the account for the jet distribution on kinetic
energy. The expected cosmic ray spectrum at the Earth is
determined with the use of a simple numerical code which takes
into account interactions of ultra-high energy protons and nuclei
with the background radiation in an expanding universe.

\end{abstract}


\end{frontmatter}

\parindent=0.5 cm

%%%%%%%%%%%%%%%%%%%%%%%%%%%%%%%%%%%%%%%%%%%%%%%%%%%%%%%%%%%%%%%%%%%%%%%%%%%%%
%% Main text
\section{Introduction}

The origin of cosmic rays with energies $E>10^{19}$ eV remains a
key problem of cosmic ray astrophysics. For a uniform distribution
of sources at the Hubble scale $c/H_{0}\approx 3\times
10^{3}h^{-1}$ Mpc, a characteristic cutoff in cosmic ray spectrum
at $5\times 10^{19}$ eV is developed (here $c$ is the velocity of
light, $H_{0}=100h$ km(s Mpc)$^{-1}$ is the Hubble parameter at
the present epoch, $h=0.7$). The cutoff arises because energetic
protons lose energy by electron-positron pair and pion production
(the GZK effect by \citet{Gr} and \citet{ZatsK}) and energetic
nuclei in addition are subject to photodisintegration
\citep{Steck}. The observed suppression of the cosmic ray flux at
energies above $\sim 5\times 10^{19}$ eV \citep{HiRes,PAO}
confirms the presence of the GZK cutoff although the suppression
due to acceleration limits in cosmic ray sources can not be
excluded. A debated topic is the value of energy $E_{c}$ in the
range $10^{17}$ eV to $10^{19}$ eV where the Galactic component
gives way to the extragalactic component in the energy spectrum of
cosmic rays observed at the Earth, see \citet{Allard,Ber09} for
discussion.

The present knowledge about the highest energy cosmic rays was
mainly acquired from the two major experiments HiRes
\citep{Abb09,Sokolsky} and Auger \citep{Abreu,PAO}. The mass
composition of the highest energy cosmic rays remains uncertain.
The interpretation of HiRes data favor the proton composition at
energies $10^{18}-5\times 10^{19}$ eV, whereas the Auger data
indicate that the cosmic ray composition is becoming heavier with
energies changing from predominantly proton at $10^{18}$ eV to
more heavy composition and probably reaching the pure Iron
composition at about $5\times 10^{19}$ eV. The mass composition
interpretation of the measured quantities depends on the assumed
hadronic model which is based on not well determined extrapolation
of the physics from lower energies. The found angular
distributions of events exceeding $6\times 10^{19}$ eV (these
particles can reach the Earth from the distances less than $\sim
100$ Mpc) were also different in these two experiments: the HiRes
data showed the isotropic distribution; the Auger data
demonstrated the correlation with AGN or objects having a spatial
distribution similar to the distribution of matter in the nearby
Universe with the largest excess from the region of the sky around
the radiogalaxy CenA. The first data on the spectrum, composition
and anisotropy from the new Telescope Array experiment support the
HiRes results \citep{Thomson}.

The list of potential sources which could give the observed flux
of the highest energy cosmic rays includes AGNs, gamma-ray bursts,
magnetars, interacting galaxies, large-scale structure formation
shocks and other objects, see e.g. review by \citet{Tor}.

It is assumed below that the AGN jets are the main extragalactic
sources of ultra-high energy cosmic rays. The kinetic energy of an
individual jet determines the maximum particle energy and the
cosmic-ray power it produces. The cases of power-law and
delta-shaped spectra of cosmic rays produced by an individual jet
are considered. The distribution of AGN jets over the kinetic
energy shapes the average source spectrum of accelerated
particles. A simple numerical code is used to calculate the
expected intensity of cosmic rays at the Earth.

\section{Transport equations for cosmic rays}

We start with a consideration of cosmic ray propagation in the
intergalactic space. The approach based on the equation for cosmic
ray density is appropriate for the purpose of the present
research.

Let us assume a uniform distribution of cosmic ray sources in the
universe. The cosmic ray number density per unit energy $N(E,t)$
and the integral number density
$N(>E,t)=\int_{E}^{\infty}dEN(E,t)$ are then also uniformly
distributed, i.e. does not depend on position $\textbf{r}$. The
total number of energetic particles in the comoving volume $V$ is
conserved and the following equation can be written down in the
absence of nuclear interactions and source contribution:

\begin{equation}
\frac{d(V(t)N(>E,t))}{dt}=\frac{\partial(V(t)N(>E,t))}{\partial
t}+ \frac{dE}{dt}\frac{\partial(V(t)N(>E,t))}{\partial E}=0
\label{conserveq}
\end{equation}
that gives

\begin{equation}
\label{elemeq} \frac{\partial N(E,t)}{\partial
t}+3H(t)N(E,t)-\frac{\partial(H(t)EN(E,t))}{\partial E}=0,
\end{equation}
where $H$ is the Hubble parameter and the expressions
$\frac{dV}{dt}=3HV$ and $\frac{dE}{dt}=-HE$ are used (the last
equation describes the energy loss rate of ultrarelativistic
particle with energy $E\approx pc$ in the expanding universe), see
\citet{Zeld}.

Additional terms that describe the continues energy losses
-$\frac{\partial((E/{\tau})N)}{\partial E}$ (if an individual
particle loses energy as $\frac{dE}{dt}=-\frac{E}{\tau(E,t)}$) and
the "catastrophic" particle losses of the form $\nu(E,t) N$ (that
describes nuclear fragmentation or radioactive decay) can be added
to the left side of Eq.~(\ref{elemeq}). The source term can be
added to its right side. Also, the redshift $z$ can be introduced
instead of $t$ variable using the relation
$\frac{dz}{dt}=-(1+z)H(z)$.Finally, the equation for nuclei with
mass number $A$ takes the form:

\begin{eqnarray}
    -H(z)(1+z)\frac{\partial}{\partial
    z}\left(\frac{F(A,\varepsilon,z)}{(1+z)^{3}}\right)- \; \nonumber \\
    \frac{\partial}{\partial\varepsilon}
    \left(\varepsilon\left(\frac{H(z)}{(1+z)^{3}}+
  \frac{1}{\tau(A,\varepsilon,z)}\right)F(A,\varepsilon,z)
  \right)+
   \nu(A,\varepsilon,z)F(A,\varepsilon,z) \nonumber \\
  = \sum_{i=1,2...}\nu(A+i\rightarrow
  A,\varepsilon,z)F(A+i,\varepsilon,z)+q(A,\varepsilon)(1+z)^{m}.\;
\end{eqnarray}

The system of equations ($3$) for all kinds of nuclei with
different $A$ should be solved simultaneously. The energy per
nucleon $\varepsilon=E/A$ is used here because it is approximately
conserved in a process of nuclear photodisintegration (analogously
to a standard practice employed in the studies of cosmic ray
nuclear spallation in the interstellar gas, see \citet{Ber90}),
$F(A,\varepsilon,z)$ is the corresponding cosmic-ray distribution
function, $q(A,\varepsilon)$ is the density of cosmic-ray sources
at the present epoch $z=0$, $m$ characterizes the source evolution
(the evolution is absent for $m=0$), $\tau(A,\varepsilon,z)$ is
the characteristic time of energy loss by the production of
$e^{-}e^{+}$ pairs and pions, $\nu(A,\varepsilon,z)$ is the
frequency of nuclear photodisintegration, the sum in the right
side of Eq. ($3$) describes the contribution of secondary nuclei
produced by the photodisintegration of heavier nuclei,
$H(z)=H_{0}((1+z)^{3}\Omega_{m}+\Omega_{\Lambda})^{1/2}$ is the
Hubble parameter in the flat expanding universe with the matter
density $\Omega_{m}(=0.3)$ and the $\Lambda$-term
$\Omega_{\Lambda}(=0.7)$.

We used Eq. ($3$) for calculations of cosmic-ray flux produced by
galaxy cluster accretion shocks \citep{Ptu09} and earlier for
calculations of proton fluxes generated by different distributions
of cosmic ray sources \citep{Ptu03}. The analogous equations were
recently used by \citep{Ber10}. The analytic study of the process
of transformation of cosmic-ray composition described in ($3$) by
the terms with frequencies $\nu$ was conducted by \citet{Hooper}.
They reproduced some mathematical results known in the theory of
nuclear transformation caused by the spallation reactions in the
interstellar gas, see \citet{Ber90}. Notice that usually the
authors prefer to build up an expression for energetic particle or
photon density in the expanding universe without explicit writing
down equations like ($3$), see e.g. basic monographs
\citet{Stecker,Ber90}. Excellent discussion on the description of
cosmic ray propagation was presented by \citet{Ber2006}.

Eq. ($3$) are valid for an arbitrary regime of cosmic ray
propagation - diffusion, rectilinear motion, or any intermediate
regime. In the case of rectilinear propagation, all functions in
Eq. ($3$) may have the direction of particle motion (determined by
the unit vector $\frac{\textbf{v}}{v}$) as an additional
independent variable that allows studying some models with
anisotropic source distribution.

The photodisintegration by the background radiation is calculated
in the present work in the approximation used by \citet{Karak}.
The corresponding cross sections are mainly taken from the works
\citet{Steck,Rach,Khan}. The spectra of the background microwave,
infrared and optical radiation are taken from \citet{Malk}. The
numerical solution of the cosmic-ray transport equations follows
the finite differences method. The variables are the redshift $z$
and $\log(E/A)$. The computation starts from some maximum $z$
($z_{max}=2$ was assumed in our calculations) and goes in the
direction of decreasing redshift till the present epoch $z=0$. At
any given $z$, the computation starts from some maximum energy per
nucleon and goes to smaller energies and to smaller atomic numbers
$A$ from Iron to Hydrogen.

It should be emphasized that an alternative Monte Carlo techniques
were used for treating the photodisintegration of ultra-high
energy cosmic rays. A representative list of references was given
by \citet{Hooper}, see also \citet{Allard08,Allard,Hooper10} where
the important results on expected cosmic ray spectra for different
source spectra and compositions were obtained. Although more
sophisticated in treating details of nuclear interactions, the
Monte Carlo techniques are more time consuming compared to the
solution of Eq.($3$) by the finite differences method and
should be used when it is required by the specific characteristics
of the problem under consideration.

The assumption of continuous source distribution is not valid when
particles lose energy at a scale less than the distance between
cosmic ray sources. The finite distance to the nearest source is
approximately taken into account in our calculations by the cutoff
of the source distribution at $z_{min}\approx 0.48H_{0}d/c << 1$,
so that $q=0$ at $z\leq z_{min}$ ($0.48d$ is the average distance
of an observer to the nearest source if the point sources arrange
the cubic lattice with the edge, the distance between sources,
equals to $d$). The statistically uniform source distribution is
assumed at larger redshifts. The rectilinear propagation of cosmic
rays from the close source is accepted here but different modes of
cosmic ray propagation can be modelled by changing the relation
between $z_{min}$ and $d$.

The imposition of an arbitrary intergalactic magnetic field does
not affect the uniform isotropic distribution of cosmic rays. The
account of cosmic-ray source granulation makes important the
presence of extragalactic magnetic fields. Magnetic effects depend
on the gyro-radius $r_{g}=10(E/10^{19}Z \textrm{eV})(B/10^{-9}
\textrm{G})^{-1}$ Mpc that determines the deflection of
ultrarelativistic particle. In principle, the presence of magnetic
field changes angular distribution of cosmic rays and cause cosmic
rays to propagate over longer distances than the straight line
distance. The weak influence of intergalactic fields on the cosmic
ray intensity was found in MHD simulations of large-scale
structures by \citet{Dolag}. Different conclusion was reached by
\citet{Sigl04}. It is clear that results critically depend on the
assumptions about the strength and structure of the intergalactic
magnetic field, see also \citep{Parizot,Kotera,Das}. Some of
magnetic field effects can by investigated in the diffusion
approximation \citep{Ber2006,BerGaziz} (additional diffusion terms
should be
 added then to the transport equations). The most straightforward but
time consuming is the direct calculation of possible trajectories
in extragalactic magnetic fields from a source to an observer. The
corresponding Monte Carlo code by \citet{Sigl04} is available on
the web. It is worth to emphasize again that the main problem in
such kind calculations is the uncertainties in the strength and
structure of extragalactic magnetic fields.

The objective of our paper, the determination of the basic overall
shapes of cosmic ray spectrum at the Earth for different source
functions produced by the AGN jets, does not require more
complicated computational procedure than solution of Eqs
($3$) with a source distribution cutoff at some $z_{min}$.

\section{Spectrum of ultra-high energy cosmic rays accelerated in AGN jets}

Simple estimates \citep{Tor,Ber90} show that from the viewpoint of
energetics the AGN jets can be the sources of ultra-high energy
cosmic rays. Different aspects of particle acceleration in AGN
jets were considered by
\citet{Biermann,Norman,Biermann2,Lemoine,Peer}, see also
references below.

To maintain the cosmic ray intensity observed in the Auger
experiment at energies above $10^{19}$ eV, the power of
extragalactic sources of the order of $3\times 10^{36}$ erg
s$^{-1}$ Mpc$^{-3}$ is required. This value increases if the
contribution of cosmic rays with smaller energies is taken into
account. At the same time the AGN jets release kinetic energy at
the level of $3\times 10^{40}$ erg s$^{-1}$ Mpc$^{-3}$ and
approximately $6$\% of this energy is contained in the jets with a
power $L_{jet}=10^{44}-10^{46}$ erg s$^{-1}$ characteristic of
FRII (Fanaroff-Riley II) radiogalaxies and radio loud quasars. We
shall use the notation FII for this population of jets. The
numerous and less powerful jets of low-luminosty AGN have power
$L_{jet}=10^{40}-10^{44}$ erg s$^{-1}$. We shall denote them as FI
sources.

Without specifying the mechanism of particle acceleration in jets,
one can use the Hillas criterion $E_{max}=Ze\beta Bl$ for the
estimate of maximum energy which the particles with charge $Ze$ in
the acceleration region of size $l$, magnetic field strength $B$,
and the velocity of magnetic field transport $u=\beta c$ can gain
(\citet{Hil84}, see also \citet{Troit}).

Let us consider the "optimistic" estimate and assume that the
energy flux of a statistically isotropic magnetic field frozen in
the jet is related to the kinetic energy flux by the relation
$L_{jet}=\beta c \frac{B^{2}}{6\pi}\pi R^{2}$. The equality
$R=l/2$ is accepted here for the jet radius; the jet velocity is
$\beta c$. As a result, the following estimate of the maximum
energy of accelerated particles can be obtained:

\begin{equation}
\label{Emax}
E_{max}=Ze\left(6\beta
c^{-1}L_{jet}\right)^{1/2}\approx2.7\times
10^{20}Z\beta^{1/2}{L_{jet,45}}^{1/2} eV,
\end{equation}
where $L_{jet,45}=L_{jet}(10^{45}\textrm{erg s}^{-1})^{-1}$, see
\citet{Lov76,Bland93,Aharonian,Wax04,Far09} and references therein
for the derivation of similar formulas.

The expression for $E_{max}$ can be also derived based on the well
studied case of the diffusive shock acceleration in young
supernova remnants. Let us consider a jet which consists of the
proton-electron plasma with the mass density $\rho$ and the power
$L_{jet}=0.5\rho u^{3} \pi R^{2}$. The cosmic rays are accelerated
at the jet termination shock and their energy density is
$w_{cr}=\eta_{cr}\rho u^{2}$, where $\eta_{cr}\approx 0.1$. The
magnetic field at the site of acceleration can reach the value of
the order $B=(4\pi \beta w_{cr})^{1/2}$ if the field is amplified
by the strong cosmic-ray streaming instability \citep{Bell}. The
maximum energy of accelerated particles is $E_{max}=Ze\beta B R$
if the Bohm diffusion near the shock is assumed (note that this
$E_{max}$ satisfies the Hillas criterium). It finally gives
$E_{max}\approx Ze\beta
\left(8\eta_{cr}c^{-1}L_{jet}\right)^{1/2}$ that is close to the
estimate~(\ref{Emax}). We shall use Eq.~(\ref{Emax}) and set
$\beta =1$ in the calculations below.

We consider two types of cosmic-ray source spectrum ejected into
the intergalactic space. The first type is a delta function
spectrum and the corresponding source power is
$q_{d}=\xi_{cr}n_{jet}L_{jet}{E_{max}}^{-1}\delta(E-E_{max})$,
where the coefficient $\xi_{cr}$ characterizes the fraction of jet
kinetic energy that goes to the accelerated particles; $n_{jet}$
is the jet number density in the intergalactic space. The second
type of sources has a power law spectrum $\propto E^{-2}$ and the
corresponding source power is
$q_{p}=\xi_{cr}n_{jet}L_{jet}E^{-2}\textrm{H}(E_{max}-E)$, where
$\textrm{H}(x)$ is the step function. (Strictly speaking, the
additional logarithmic normalization factor
$(\ln(E_{max}/E_{min}))^{-1}$ should by included in the last
expression for $q_{p}$. We omit it because of uncertain value of
the minimal energy $E_{min}$ for particles ejected from the
accelerator.)

It can be recalled as an example that both spectrum shapes of
ejected particles arise in the consideration of diffusive shock
acceleration in supernova remnants, see \citet{PtZir05}. The
runaway particles have close to the delta function energy spectrum
$\sim\delta(E-E_{max})$ where $E_{max}$ is the maximum energy of
accelerated particles that is achieved at the given stage of a
supernova shock evolution. The energetic particles that remains
confined inside the remnant may have close to a power law spectrum
and leave out into the interstellar medium at some stage of SNR
evolution when the shock breaks up. It should be stressed that we
do not assume that two discussed injection spectra work at the
same time and analyze them separately.

\begin{figure}[tb]
\begin{center}
\includegraphics[width=10.0cm]{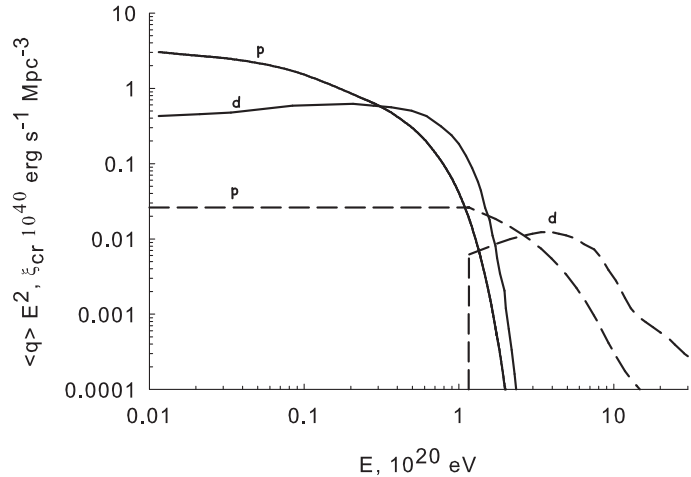}
\end{center}
\caption{Calculated average source spectra of jet populations FI
(solid lines) and FII (dash line) for delta-function (\emph{d})
and power law $E^{-2}$ (\emph{p}) cosmic ray spectra generated by
individual jets. Eq.~(\ref{Emax}) was used for calculations of
$E_{max}(L_{jet})$} \label{source}
\end{figure}

The source functions $q_{d}(E)$ and $q_{p}(E)$ should be averaged
over the distribution of jet luminosity $n_{jet}(L_{jet})$ to
obtain the average source function of extragalactic cosmic rays.
The results are shown in Fig.~(\ref{source}) for the functions
$n_{jet}(L_{jet})$ presented by \cite{Koer08} in their Figure
($8$) for the kinetic luminosity function of jets. Four source
functions in Fig.~(\ref{source}) correspond to the combination of
two populations of jets, FI and FII, and two types of jet spectra,
\emph{d} and \emph{p}. These four source functions are considered
here as representing four different scenarios of cosmic ray
acceleration in jets.

\begin{figure}[tb]
\begin{center}
\includegraphics[width=10.0cm]{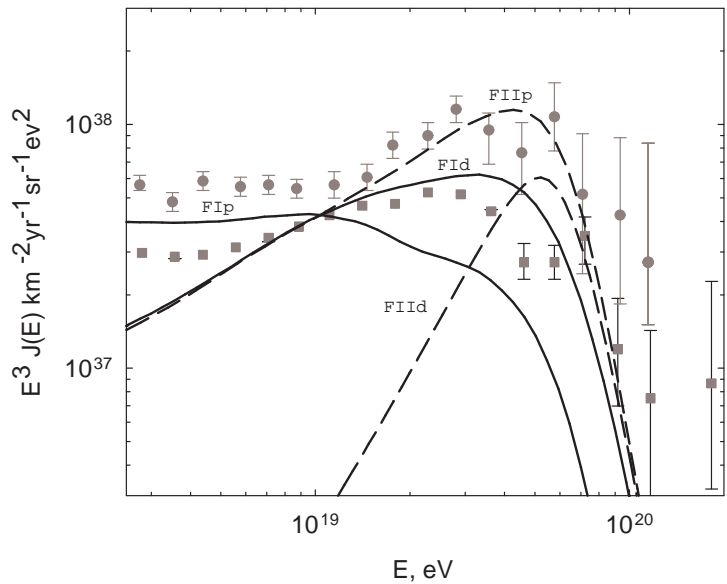}
\end{center}
\caption{Calculated spectra of extragalactic cosmic rays for
sources $<q_{d}>$ and $<q_{p}>$ averaged over the AGN jet
population FI (solid lines FId and FIp) and FII (dash lines FIId
and FIIp). The spectra (except FIId) are normalized to the
intensity observed at $10^{19}$ eV in the Auger experiment. Data
are from HiRes experiment \citep{Abb09} (circles) and Auger
experiment \citep{PAO} (squares).}
\label{spectr}
\end{figure}

We solved numerically the set of transport Eqs ($3$) for the
described types of the average source spectrum.
Fig.~(\ref{spectr}) illustrates the results. It was assumed that
the source chemical composition coincides with the composition of
Galactic cosmic ray sources. The spectra were normalized at
$10^{19}$ eV to the observed by Auger intensity. It requires very
different efficiency of particle acceleration
$\xi_{cr,FII}/\xi_{cr,FI}\sim 20$ in the FII and FI jets. These
efficiencies are $\xi_{cr,FII}\sim 0.1$ and $xi_{cr,FI}\sim 0.005$
if the source spectra are extrapolated down to $1$ GeV. No
cosmological evolution was assumed in our calculations ($m=0$).
The evolution is different for different morphological types of
AGN, see e.g. \citet{Ber2006}, but it does not significantly
affect the calculated spectra at energies $>3\times10^{18}$ eV
since these particles may come from the distances not larger than
about $2\times10^{3}$ Mpc.

Of four spectra shown in Fig.~(\ref{spectr}), two reproduce
cosmic ray observations with reasonable accuracy. They correspond
to the AGN jet population FI with delta-function source spectra
and the AGN jet population FII with power-law spectra $E^{-2}$.

The finite distance $z_{min}$ to the closest to an observer source
was taken into account as discussed in the preceding Section. This
distance is a function of particle energy and charge and is
different for the source populations FI and FII. The absence of
sources at distances $<90$ Mpc in our model resulted in a steeply
sloping down spectrum of cosmic rays at the highest energies
$10^{20}$ eV for the FII source distribution.

The dependence $n_{jet}(L_{jet})$ together with Eq.~(\ref{Emax})
 for $E_{max}(L_{jet})$ leads to the dependence of
cosmic-ray source number density on particle energy $n_{s}(E)$.
For FI population of sources with delta shaped spectra, the source
density is $n_{s}=10^{-4}$ Mpc$^{-3}$ at $E=6\times10^{19}$ eV and
$n_{s}=2\times10^{-3}$ Mpc$^{-3}$ at $E=10^{19}$ eV that coincides
with the results of \citet{Takami} derived from the analysis of
cosmic-ray arrival direction distribution.

\begin{figure}[tb]
\begin{center}
\includegraphics[width=10.0cm]{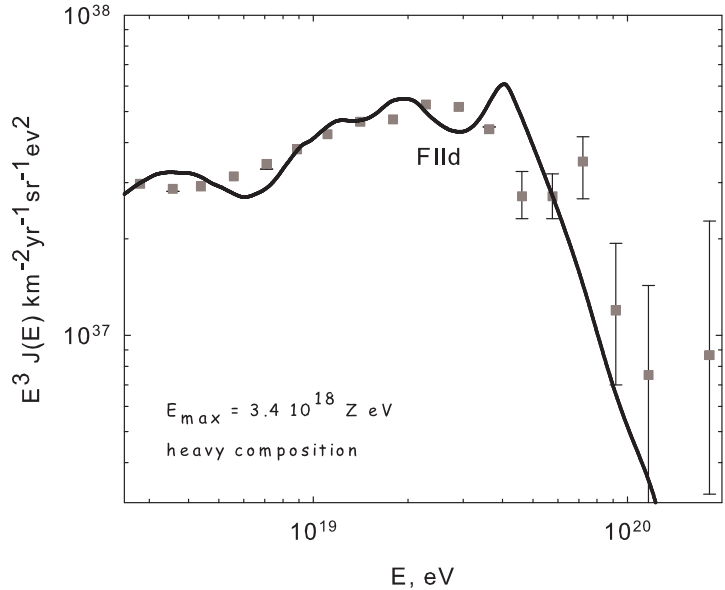}
\end{center}
\caption{Calculated spectrum of extragalactic cosmic rays for
sources $<q_{d}>$ averaged over the AGN jet population FIId with
heavy composition; $E_{max}$ is decreased by $80$ compared to
Eq.~(\ref{Emax}). Data are from Auger experiment \citep{PAO}
(squares).} \label{heavy}
\end{figure}

Protons dominate in the calculated composition of cosmic rays that
is in strong disagreement with the Auger data. To get out of a
difficulty one can take anomalously high abundance of heavy nuclei
and reduce the maximum particle energy at the source
\citep{Allard,Hooper10}. We reserve the consideration of this
issue for a later paper and show in Fig.~(\ref{heavy}) only one
example of calculations where the shape of source spectrum
corresponds to the FIId sources with the maximum particle energy
$E_{max}$ decreased by a factor of $80$ compared to the
"optimistic" estimate (\ref{Emax}) (this relieves the extreme
assumptions used in the derivations of Eq.~(\ref{Emax})) and
the Iron abundance at the source comprises $1/3$ of all nuclei.
The calculated cosmic-ray spectrum only roughly reproduce the
observed spectrum. The cosmic-ray composition can be characterized
by the mean logarithmic atomic number $<\ln (A)>$. Its calculated
value rises approximately linearly from about $0.25$ at $5\times
10^{18}$ eV to $3.7$ at $5\times 10^{19}$ eV in a qualitative
agreement with the Auger data on the shower maximum dependence on
energy.

\section{Conclusions}

It is believed that the particle acceleration by jets in active
galactic nuclei is the most efficient source of cosmic rays with
the highest energies, $E > 10^{19}$ eV. In the present work we
distinguish two populations of jets: FI produced by the low
luminosity AGN populations with jet power $2\times 10^{40}$ to
$3\times 10^{44}$ erg s$^{=1}$, and FII produced by high
luminosity AGN with jet power larger than $2\times 10^{44}$. The
corresponding jet distributions on power were given by
\citet{Koer08}.

The typical power law spectrum of nonthermal jet radiation implies
the power law particle spectrum of the form close to $E^{-2}$. One
may expect that the spectrum of particles released into the
extragalactic space, the source spectrum of extragalactic cosmic
rays, has the same shape. Another possibility is that accelerated
particles remains confined inside the source and only particles
with maximum energies run away into intergalactic space so that
the source spectrum is of a delta-shaped form. The average source
spectrum of extragalactic cosmic rays is determined as the
convolution of one of these source functions of an individual jet
with the jet distribution on power. Based on the Hillas criterion,
we accepted an optimistic estimate for the maximum energy of
accelerated particles (\ref{Emax}) with its characteristic scaling
$E_{max}\propto L_{jet}^{1/2}$ and used it in our calculations.

The computations of cosmic ray propagation in the expanding
Universe filled with the background electromagnetic radiation were
fulfilled by the use of a simple numerical code which solves the
system of coupled transport Eqs~($3$) for energetic protons
and nuclei from He to Fe. The calculations were made under the
approximation of continuous energy losses by $e^{-},\ e^{+}$ and
pion production and the "catastrophic" losses through
photodisintegration and corresponding production of secondary
nuclei.

The results of our calculations are illustrated in
Fig.~(\ref{spectr}). The observed spectrum of ultra-high energy
cosmic rays can in principle be explained in the frameworks of two
scenarios - the acceleration by the FI sources with individual jet
spectra of the delta-shaped form or the acceleration by the FII
sources with individual jet spectra close to $E^{-2}$ form. The
transition from Galactic to extragalactic component in the
observed at the Earth spectrum occurs at about $(3...5)\times
10^{18}$eV.

The calculated spectra are normalized to the observed by Auger
intensity at $10^{19}$ eV. It requires very different efficiencies
of transformation of the jet kinetic energy to the energy of
cosmic rays in the FII and FI sources:
$\xi_{cr,FII}/\xi_{cr,FI}\sim 20$. It was assumed that the
elemental composition of accelerated particles is the same as in
the Galactic cosmic-ray sources. This results in the predominantly
proton composition of ultra high energy extragalactic cosmic rays
that is compatible with the HiRes data but not with the Auger
data. One needs to significantly increase the abundance of heavy
nuclei at the source and drastically decrease the value of
$E_{max}$ to fit the Auger data \citep{Allard,Hooper10}. This
procedure was discussed at the end of Section $3$ and illustrated
by Fig.~\ref{heavy}.

\bigskip
The work was supported by the Russian Foundation for Basic
Research grant 10-02-00110a.

\end{document}